\documentclass{article}
\usepackage{arxiv}

\usepackage[utf8]{inputenc} 
\usepackage[T1]{fontenc}    
\usepackage{amssymb, amsmath}
\usepackage{mathabx}	

\usepackage[boxruled,linesnumbered,commentsnumbered]{algorithm2e}

\SetCommentSty{mycommfont}

\usepackage{verbatim}
\usepackage{makeidx}         
\usepackage{graphicx}
\usepackage{epstopdf}
\usepackage{setspace}
\usepackage{url}

\usepackage{cite}
\usepackage{amsmath,amssymb,amsfonts}
\usepackage{mathabx}	
\usepackage{algorithmic}
\usepackage{comment}

\usepackage{subfigure}
\usepackage{balance}
\usepackage{textcomp}
\usepackage{xcolor}
\def\BibTeX{{\rm B\kern-.05em{\sc i\kern-.025em b}\kern-.08em
    T\kern-.1667em\lower.7ex\hbox{E}\kern-.125emX}}

\usepackage{indentfirst}

\usepackage{url}

\usepackage{color}

\title{You Shall not Pass: Avoiding Spurious Paths in Shortest-Path Based Centralities in Multidimensional Complex Networks}

\author{
Klaus Wehmuth; Artur Ziviani\\
Computer Science Dept\\ National Laboratory for Scientific Computing (LNCC)\\25651-075, Petr\'{o}polis, RJ, Brazil\\
\texttt{klaus@lncc.br; ziviani@lncc.br}\\
\And
Leonardo C. Costa; Alex B. Vieira\\
Computer Science Dept\\ Universidade Federal de Juiz de Fora (UFJF)\\36036-330, Juiz de Fora, MG, Brazil\\
\texttt{leonardo.chinelate@gmail.com; alex.borges@ufjf.edu.br}\\
\And
Ana Paula Couto da Silva\\
Computer Science Dept\\ Universidade Federal de Minas Gerais (UFMG)\\31270-010, Belo Horizonte, MG, Brazil\\
\texttt{ana.coutosilva@dcc.ufmg.br}
}

\begin{document}
\maketitle

\begin{abstract}
In complex network analysis, centralities based on shortest paths, such as betweenness and closeness, are widely used. More recently, many complex systems are being represented by time-varying, multilayer, and time-varying multilayer networks, i.e. multidimensional (or high order) networks. Nevertheless, it is well-known that the aggregation process may create spurious paths on the aggregated view of such multidimensional (high order) networks. Consequently, these spurious paths may then cause shortest-path based centrality metrics to produce incorrect results, thus undermining the network centrality analysis. In this context, we propose a method able to avoid taking into account spurious paths when computing centralities based on shortest paths in multidimensional (or high order) networks. Our method is based on MultiAspect Graphs~(MAG) to represent the multidimensional networks and we show that well-known centrality algorithms can be straightforwardly adapted to the MAG environment. Moreover, we show that, by using this MAG representation, pitfalls usually associated with spurious paths resulting from aggregation in multidimensional networks can be avoided at the time of the aggregation process. As a result, shortest-path based centralities are assured to be computed correctly for multidimensional networks, without taking into account spurious paths that could otherwise lead to incorrect results. We also present a case study that shows the impact of spurious paths in the computing of shortest paths and consequently of shortest-path based centralities, such as betweenness and closeness, thus illustrating the importance of this contribution.
\end{abstract}



\section{Introduction}
\label{sec:intro}

Centrality is a key concept for complex network analysis~\cite{ Bonacich1987, Freeman1979}. Typically seen as a ranking of the relative importance of vertices or edges in a complex network, many distinct centrality definitions have been proposed over the last decades for different purposes in several fields~\cite{Brin1998, Holme2012,Jeong2001,Pan2011,Sole-Ribalta2014}.  More recently, many complex systems are being represented by time-varying, multilayer, and time-varying multilayer networks, i.e. multidimensional~(or high order) networks. Traditionally, multidimensional networks were referred to as networks for which there are more than one edge connecting two vertices, i.e. a multigraph, or as
networks where the dynamics of nodes are with multiple dimensions, such as having their location coordinates in 3D space.\footnote{We remark that all considerations and contributions of this paper about connectivity and avoidance of spurious paths remain valid if nodes may have multiple edges between them (a multigraph) or may have location coordinates in 3D space, for instance.}
However, with the emergence of time-varying and multilayer networks, the definition of multidimensional networks evolved to support these newer types of higher order networks. Berlinger et al.~\cite{Berlingerio2011} state that dimensions in network data can be either explicit or implicit. In the first case the dimensions directly reflect the various interactions in reality; in the second case, the dimensions are defined by the analyst to reflect different interesting qualities of the interactions, which can be inferred from the available data, in structures as multilayer or multiplex networks. Therefore, inline with this interpretation of several related works~\cite{Johnson1995,Zhao2011,Berlingerio:2013,Kivela2014,Wehmuth2015b}, in this paper, we use the term multidimensional network to refer to higher-order complex networks that involve multiple aspects or features (i.e. dimensions), such as time instants, layers, and so on. More specifically, in this paper,  we focus on shortest-path based centralities, such as betweenness and closeness, in multidimensional networks.

We represent multidimensional networks by means of using MultiAspect Graphs~(MAGs)~\cite{Wehmuth2016, Wehmuth2017}, which are a recently introduced abstraction able to represent time-varying, multilayer, combined time-varying and multilayer, or even higher order graphs. In a MAG, each independent structure (i.e., a dimension) of a high order network is represented by an aspect. Therefore, in a MAG, aspects are used to represent the vertices set, layer set, time instants set, and so on.  MAGs are shown to be isomorphic to an object composed of a directed graph and one companion tuple. Furthermore, MAGs are closely related to directed graphs~\cite{Wehmuth2016}. 
Algebraic representations and basic algorithms for MAGs are investigated in~\cite{Wehmuth2017}. 
These previous works pave the way for the MAG centrality notions investigated here. In particular, we focus on shortest-path based centralities, since they are widely applied in complex network analysis and relatively straightforward to interpret. 

\begin{figure}[htbp]
        \centering
        \includegraphics[width=0.5\textwidth]{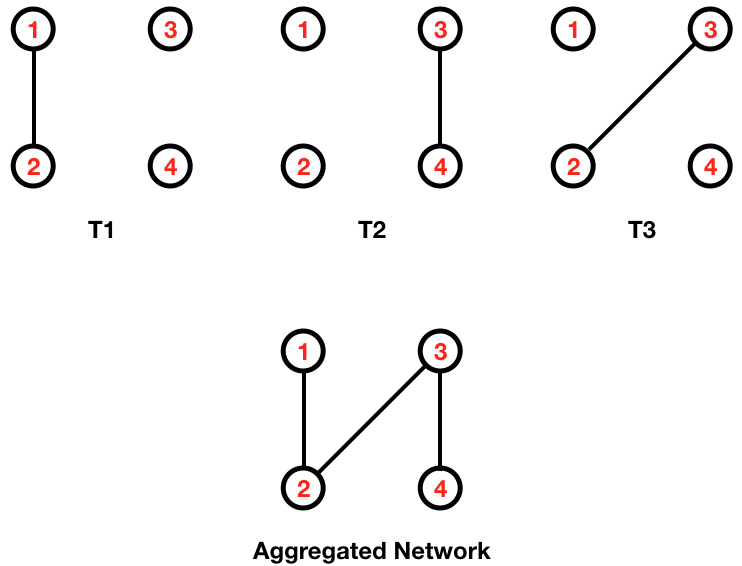}
        \caption{Time-Varying Graph and its aggregated form.}
        \label{fig:TVG_Agg}
\end{figure}

In order to assess node centralities in multidimensional networks, it is common to aggregate one or more aspects of the network in a dimensionality reduction process. Nevertheless,  it is well-known that the aggregation process may create spurious paths on the aggregated view of such multidimensional networks~\cite{Pan2011, Nicosia2012b, Ribeiro2013, Berlingerio:2013, Kivela2014, DeDomenico2016}.
Spurious paths emerge as additional paths that actually did not exist in the original network. As a consequence, this may generate shortest paths that also did not exist in the original networks. The presence of these artifacts on the aggregated network causes the shortest-path based centrality measures to include spurious shortest paths on their computation, leading to results that are not consistent with the original network and undermining the network centrality-based analysis of the multidimensional network. Figure~\ref{fig:TVG_Agg} presents a small illustrative example of this issue. In this case, T1, T2, and T3 represent three instants of a time-varying graph (TVG). It can be seen that in this TVG there is no path from vertex 1 at time T1 to vertex 4 at any time since the edge $\{3,4\}$ occurs at time~T2, before edge $\{2,3\}$ at time T3. However, in the aggregated network there is a path from vertex~1 to vertex~4. Often  such spurious paths in aggregated network views are overlooked, thus potentially misleading the centrality analysis, or it becomes a cumbersome (and not scalable) process to disregard them in computing shortest paths at the aggregated view. This motivates the investigation of ways to avoid these spurious paths at the aggregation process.

In this paper, we propose a method\footnote{This is an extended version of the shorter paper~\cite{Wehmuth-ladc2018} that showed only the theoretical part of the work.} 
able to avoid taking into account spurious paths when aggregating dimensions and, as a consequence, also when later computing centralities based on shortest paths in multidimensional networks. To the best of our knowledge, this is the first method to avoid the generation of spurious paths at the aggregation process in multidimensional networks. Further, we present a formalization of well-known centralities in the MAG environment, assuring that the subject of shortest-paths centralities in MAGs is well defined to provide the support proposed solution. The presence of aggregation artifacts on the aggregated network causes the centrality algorithms based on shortest paths to include these shortest \emph{spurious} paths into the centrality calculation, leading to results that are not consistent with the original network. Then, as the main contribution of this work, we prove that, by using the proposed method, pitfalls usually associated with aggregation can be avoided in multidimensional networks. 
As a result, path-based centralities are assured to be computed correctly without taking into account spurious paths that could lead to incorrect results. Finally, we present a case study that shows the impact of spurious paths in the computing of shortest paths and consequently of shortest-path based centralities, such as betweenness and closeness. This also demonstrates the importance of this contribution by comparing the difference between results from traditional method and results from our proposed method that avoids spurious paths.

This paper is organized as follows. Section~\ref{sec:def_mag} presents a brief description of MAGs. Section~\ref{sec:magsub} defines sub-determination, illustrates how a spurious path may emerge from it, and present the proposed method that enables the  avoidance of spurious paths in the sub-determination process, which constitutes the main contribution of the paper. Section~\ref{sec:cent_mag} formalizes shortest-path based centralities and their sub-determined form on the MAG environment. Section~\ref{sec:Algs} discusses the implementation of algorithms for evaluating shortest-path centralities on MAGs and presents an example of a betweenness centrality algorithm in both sub-determined and non-sub-determined forms. Section~\ref{sec:app} presents a case study comparing results obtained by the traditional method to results obtained by our proposed method that avoids spurious paths. Finally, Section~\ref{sec:rem} presents final remarks and concludes the paper.

\section{Background on MultiAspect Graphs (MAGs)}
\label{sec:def_mag}
We define a MAG as $H = (A, E)$, where $E$ is a set of edges and $A$ is a finite list of sets, each of which is called an \emph{aspect}. Each aspect $\sigma \in A$ is a finite set, and the number of aspects $p = |A|$ is called the order of $H$. Each edge $e \in E$ is a tuple with $2 \times p $ elements. All edges are constructed so that they are of the form $(a_1,\dots,a_p,  b_1,\dots,b_p)$, where $a_1, b_1$ are elements of the first aspect of $H$, $a_2,b_2$ are elements of the second aspect of $H$, and so on, until $a_p, b_p$ which are elements of the $p$-th aspect of $H$.

As a matter of notation, we say that $A(H)$ is the aspect list of $H$ and $E(H)$ is the edge set of~$H$. Further, $A(H)[n]$ is the $n$-th aspect in $A(H)$,  $|A(H)[n]| = \tau_n$ is the number of elements in $A(H)[n]$, and $p = |A(H)|$ is the order of $H$. 
Further, we define the set
\begin{equation}
\label{eq:VH}
\mathbb{V}(H) = \bigtimes_{n=1}^p A(H)[n],
\end{equation}
which is the Cartesian product of all the $p$ aspects of the MAG $H$. We call the elements $\mathbf{v} \in \mathbb{V}(H)$ \emph{composite vertices}.
Note that each composite vertex $\mathbf{v} \in \mathbb{V}(H)$ has the form $(a_1,\dots,a_p)$. Therefore, there is a close relation between MAG edges and pairs of composite vertices, since $(a_1,\dots,a_p,  b_1,\dots,b_p) \sim (\mathbf{v}, \mathbf{u}) = ((a_1,\dots,a_p),$ $(b_1,\dots,b_p))$, so that $\mathbf{v} = (a_1,\dots,a_p)$ and $\mathbf{u} = (b_1,\dots,b_p)$.
From this relation between MAG edges and pairs of composite vertices, it is possible to build a directed graph of composite vertices, which is shown in~\cite{Wehmuth2016} to be isomorphic to the MAG. 

As a consequence of the isomorphism between a MAG and a directed graph, we then define the function
\begin{align}
\label{func:iso}
g: (A(H), E(H))& \to ( \mathbb{V}(H),\mathbb{V}(H) \bigtimes \mathbb{V}(H)) \\
H  & \mapsto  (\mathbb{V}(H),  \psi(E(H)), \notag
\end{align}
which maps every MAG~$H$ to its isomorphic directed graph $g(H)$.
Further, we define the set of functions
\begin{align}
\label{eq:pi_i}
\pi_i:\mathbb{V}(H) & \to A(H)[i] \\
(a_1,a_2,\dots,a_p) & \mapsto a_i, \notag
\end{align}
which extracts the $n$-th element of a composite vertex tuple.

Note that, by the definition of $g(H)$ in Expression~\ref{func:iso}, the vertices of the directed graph $g(H)$ are tuples with $p$ elements. It is also possible, if desired, to have a more traditional graph representation where the vertices are simply points of a set with no additional meaning. In this case, the directed graph has to be complemented by a \emph{companion tuple}, which allows the association of each vertex with its original MAG tuple.
This companion tuple has $p$ elements, where each element $0 < i \leq p$ is given as $|A(H)[i]|$, the number of components of the $i$-th aspect. Further details regarding the construction and usage of the companion tuple of a MAG can be found in~\cite{Wehmuth2017}.

Since every MAG $H$ is isomorphic to a directed graph, it follows that shortest paths between composite vertices are equivalent to shortest paths between vertices of a directed graph. Therefore, algorithms based on shortest paths between composite vertices can be constructed as traditional directed graph algorithms. In particular, algorithms for centralities based on shortest paths, like for instance, closeness and betweenness centralities, can be computed for composite vertices on a MAG with the same algorithms used for computing them in directed graphs.

In the case of sub-determined vertices on a MAG, however, the traditional algorithms for directed graphs may lead to problems, if applied directly. As shown in~\cite{Wehmuth2016,Wehmuth2017}, shortest paths and distances between sub-determined vertices do not have the same properties of shortest paths and distances on directed graphs.

In MAGs, similarly to traditional graphs, distances can be defined in different ways depending on the application. In this sense, one common way of defining distances between composite vertices in MAGs is the number of hops of a shortest path between these vertices. Edge attributes (weights) can be used to assign distances (e.g. metric distances) to an edge. Moreover, any MAG that has time instants as an aspect can be seen as a time-varying MAG. Hence, the time length of an edge is determined by the difference between the two time values on that edge. Therefore, time distances can be inferred directly from the edge composition. As a consequence, as with traditional directed graphs, algorithms based on shortest paths can be adapted for use with any other distance definition.


\section{Sub-determination and \\the avoidance of spurious paths}
\label{sec:magsub}

In this section, we define the sub-determination process and we illustrate how a spurious path may emerge from it. Finally, we formally present the proposed method that enables the avoidance of spurious paths in the sub-determination process.

\subsection{Sub-determination}

The sub-determination process can be seen as a partition of the original MAG, where composite vertices that are sub-determined to the same sub-determined vertex are on the same equivalence class.
In other words, a MAG sub-determination is a generalization of the aggregation applied to multilayer or time-varying graphs, in which all layers or time instants can be aggregated, resulting in a traditional directed graph.  

Since a MAG can have more than 2~aspects, a sub-determination can be performed in more ways than the aggregation.
For instance, given an arbitrary MAG~$H$, a sub-determination is constructed from a non-empty proper sublist of the aspects of $H$. Since this sublist excludes the empty sublist and the full aspect list of $H$, in a MAG with $p$ aspects there are $2^p - 2$ possible non-empty proper aspect sublists, each one of them originating a distinct sub-determination. Therefore, in a MAG~$H$ with $p$ aspects we identify the possible sub-determination by a positive integer $\zeta$, where $1 \leq \zeta \leq 2^p - 2$. 

A tuple with $p$ elements either $0$ or $1$ can be used as a set indicator to represent each possible sub-determination. An element with value $1$ in the $i$-th position of the tuple indicates that the $i$-th aspect is present in the sub-determination while a $0$ indicates that the $i$-th aspect is not present in the sub-determination. We remark that this tuple is equivalent to the binary representation of the integer $\zeta$. 
As a consequence, we may use the symbol $\zeta$ to represent the sub-determination both as an integer number or its equivalent tuple. 
Note that we adopt the notation $\zeta$ to denote any given sub-determination, whereas for a specific sub-determination we use a subscript to characterize the specific tuple associated with the sub-determination. For example, a sub-determination generated by the tuple $\mathbf{Y} = [1,0,1]$ is represented as $\zeta_\mathbf{Y}$. 

For a given MAG~$H$ and a sub-determination $\zeta$, we have a sublist $A_\zeta(H) \subset A(H)$ such that $p_\zeta = |A_\zeta(H)|$ is the number of aspects in the sub-determination. From this, we define the set of composite sub-determined vertices of $H$ as
\begin{equation}
\label{eq:subvtx}
\mathbb{V}_\zeta(H) = \bigtimes_{n-1}^{p_\zeta}  A_\zeta(H)[n],
\end{equation}
i.e., the Cartesian product of all the aspects in the sub-determination $\zeta$. Further, we also define the function
\begin{align}
\label{func:subS}
S_\zeta: \mathbb{V}(H) & \to \mathbb{V}_\zeta(H) \\
(a_1, a_2, \dots, a_p)  & \mapsto  (a_{\zeta_1}, a_{\zeta_2}, \dots, a_{\zeta_{p_\zeta}}), \notag
\end{align}
which takes each composite vertex of $H$ to its sub-determined form.

Even though aggregation (or sub-determination for the case of MAGs) may cause additional paths to be present on the aggregated network, we show that it  is possible to obtain a sub-determined Breadth-First Search~(BFS) in which the additional paths potentially created by the sub-determination/aggregation process are not considered. The proof of this claim can be found in~\cite{Wehmuth2017}. However, in order to make this work self-contained, we present a brief motivation of the proof, providing a deeper understanding of how the issue of the spurious path can be avoided. 

The construction of sub-determined algorithms relies on the use of functions to aggregate/reduce results according to the applied sub-determination. In some cases, this function can be as simple as just summing up values obtained in composite vertices, which are reduced to the same sub-determined vertex. It follows that such summation can be done by matrix multiplication. Given a MAG~$H$ and a sub-determination $\zeta$, the sub-determination matrix $\mathbf{M}_\zeta(H)$ $\in \mathbb{R}^{n_\zeta \times n}$ is a rectangular matrix, where $n = |\mathbb{V}(H)|$ is the number of composite vertices of $H$ and $n_\zeta = |\mathbb{V}_\zeta(H)| $ is the number of composite vertices of the sub-determination~$\zeta$ applied to the MAG~$H$. Since a sub-determination is a~(proper) subset of the aspects of a MAG, it follows that $n_\zeta | n$, i.e. the number of composite vertices of a MAG, is a multiple of the number of composite vertices in any of its sub-determinations. Further, $\mathbf{M}_\zeta(H)$ has the property of having exactly one non-zero entry in each column; and the position of this entry is determined by the numerical value of the sub-determined composite vertex.

Algorithm~\ref{alg:M_zeta} shows the construction of the sub-determination matrix $\mathbf{M}_\zeta(H)$ for a given MAG~$H$ and sub-determination $\zeta$. As Algorithm~~\ref{alg:M_zeta} is actually a classic BFS, it has time complexity $O(n+n_\zeta)$ and space complexity~$O(n)$.
The function $D$ takes a composite vertex to its numerical representation.

\begin{algorithm}[ht]
\DontPrintSemicolon
	\SetKwData{Left}{left}\SetKwData{This}{this}\SetKwData{Up}{up} 
	\SetKwFunction{Union}{Union}\SetKwFunction{FindCompress}{FindCompress} 
	\SetKwInOut{Input}{input}\SetKwInOut{Output}{output}
	\SetKwFunction{algo}{algo}
        \SetKwProg{myalg}{SubDetMatrix($\tau(H)$, $\zeta$)}{}{}
	
	\Input{$\tau(H) \text{ and } \zeta$}
	\Output{$\mathbf{M}_\zeta(H)$}
	\BlankLine
	\myalg{} {
		$T_\zeta = SubCompTuple(\tau(H), \zeta)$ \tcp*{$\zeta$ sub-determined companion tuple}
		$n \gets |\mathbb{V}(H)|$ \;
		$n_\zeta \gets|\mathbb{V}_\zeta(H)|$ \;
		$\mathbf{M}_\zeta(H) \gets n_\zeta \times n$ \tcp*{ sparse matrix} 
		\For {$j \gets 1$ \textbf{to} $n$} {
			$\mathbf{u} \gets D^{-1}(j, \tau(H))$ \tcp*{numeric tuple form of j}
			$i \gets D(\mathbf{u},T_\zeta)$ \tcp*{sub-determined representation}
			$\mathbf{M}_\zeta(H)[i,j] \gets 1$ \;
		}
	}
	\Return{$\mathbf{M}_\zeta(H$)}\;
\caption{Construction of $\mathbf{M}_\zeta$.}
\label{alg:M_zeta}
\end{algorithm}

\begin{figure*}[!ht]
	\centering
		\subfigure[MAG $R$]{\includegraphics[width=2.0in]{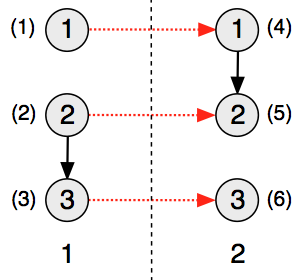} \label{fig_a}}
		\hfil
		\subfigure[sub-determined $R$]{\includegraphics[width=1.5in]{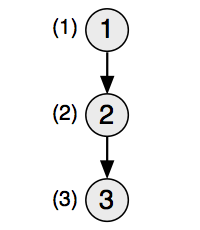} \label{fig_b}}
	\caption{Order~2 MAG $R$ with 2 time instants and its sub-determined form (in time).}
    \label{fig_mag}
\end{figure*}

From Algorithm~\ref{alg:M_zeta}, it can be seen that the sub-determination matrix $\mathbf{M_{\zeta_T}}(R)$, used to do the sub-determination of the MAG~$R$ shown in Figure~\ref{fig_mag}, has the form depicted in Equation~\ref{eq:MzR}. Here, the sub-determination applied is given by the tuple $\mathbf{T} = [1,0]$, in this case sub-determining the time aspect on the MAG~$R$.

\begin{equation}
\label{eq:MzR}
\mathbf{M_{\zeta_T}}(R) =   \left[ 
\footnotesize
\begin{array}{rrrrrr}
\mathbf{1}& 0 & 0 & \mathbf{1} & 0 & 0\\
0 & \mathbf{1} & 0 & 0 & \mathbf{1} & 0\\
0 & 0 & \mathbf{1} & 0 & 0 & \mathbf{1}\\
\end{array} \right]
\end{equation}

\subsection{An example of the emergence of a spurious path}

Given the adjacency matrix $\mathbf{J}(H)$ of an arbitrary MAG~$H$, and given a sub-determination matrix $\mathbf{M}_\zeta(H)$ that represents the desired sub-determination, the adjacency matrix of the sub-determined MAG~$H_\zeta$ is obtained by 
\begin{equation}
\label{eq:sub}
\mathbf{J_\zeta}(H) = \mathbf{M}_\zeta(H) \ \mathbf{J}(H) \ \mathbf{M}_\zeta(H)^T.
\end{equation}
The adjacency matrix $\mathbf{J}(R)$ of the MAG $R$ in Figure~\ref{fig_a} is then given by
\begin{equation}
\label{eq:JR}
\mathbf{J}(R) =   \left[ 
\footnotesize
\begin{array}{rrrrrr}
0 & 0 & 0 & \mathbf{1} & 0 & 0\\
0 & 0 & \mathbf{1} & 0 & \mathbf{1} & 0\\
0 & 0 & 0 & 0 & 0 & \mathbf{1}\\
0 & 0 & 0 & 0 & \mathbf{1} & 0\\
0 & 0 & 0 & 0 & 0 & 0\\
0 & 0 & 0 & 0 & 0 & 0\\
\end{array} \right].
\end{equation}

Therefore, it follows that the adjacency matrix $\mathbf{J_{\zeta_T}}(R)$ of the sub-determined MAG~$R$, shown in Figure~\ref{fig_b}, is given by
\begin{equation}
\label{eq:JzR}
\mathbf{J_{\zeta_T}}(R) = \mathbf{M_{\zeta_T}}(R) \ \mathbf{J}(R) \ \mathbf{M_{\zeta_T}}(R)^T,
\end{equation}
so that
\begin{equation}
\label{eq:MJzRm}
\mathbf{J_{\zeta_T}}(R) = \left[ 
\footnotesize
\begin{array}{rrr}
\mathbf{1} & \mathbf{1} & 0 \\
0 & \mathbf{1} & \mathbf{1} \\
0 & 0 & \mathbf{1} \\
\end{array} \right].
\end{equation}
Due to the properties of matrix multiplication, it follows that the result of the algebraic sub-determination of a MAG results in a multi-graph with self-loops. Since we usually consider simple graphs, we may replace the main diagonal by a zero diagonal and (if necessary) substitute any non-zero entry by~1. In this way, the adjacency matrix $\mathbf{\hat{J}_{\zeta_T}}(R)$
corresponding to Figure~\ref{fig_b} is given by
\begin{equation}
\label{eq:JzRm}
\mathbf{\hat{J}_{\zeta_T}}(R) = \left[ 
\footnotesize
\begin{array}{rrr}
0 & \mathbf{1} & 0 \\
0 & 0 & \mathbf{1} \\
0 & 0 & 0 \\
\end{array} \right].
\end{equation}
Note that the spurious path from vertex $2$ to vertex $3$ in Figure~\ref{fig_b} is present in the adjacency matrix $\mathbf{\hat{J}_{\zeta_T}}(R)$, as expected.

\subsection{The avoidance of spurious paths}

We now show that given an arbitrary MAG~$H$, its adjacency matrix $\mathbf{J}(H)$, and a sub-determination $\zeta$, it is possible to obtain a BFS sub-determined by $\zeta$, in which no spurious path is considered. The BFS is closely related to matrix multiplication. This stems from the well-known property of the powers of the adjacency matrix, in which the $(i,j)$ entry of the $n$-th power of the adjacency matrix shows the number of existing walks of length $n$ from vertex $i$ to vertex $j$~\cite{Kepner2011}. From this, we could think that for a given MAG~$H$, the series
\begin{equation}
\label{eq:bfs_non_conv}
\mathbf{B} = \sum_{i=0}^{\infty} \mathbf{J}(H)^i = \mathbf{I} + \mathbf{J}(H) + \mathbf{J}(H)^2 +  \mathbf{J}(H)^3 + \mathbf{J}(H)^4 +\dots
\end{equation}
would produce a matrix $\mathbf{B}$, such that the entry $\mathbf{B}_{i,j}$ indicates the number of walks of any length from vertex $i$ to vertex $j$. This is indeed the case when $H$ happens to be an acyclic MAG, making $\mathbf{J}(H)$ a nilpotent matrix. The existence of cycles in $H$ makes that, for some vertices, there will exist walks of arbitrary length connecting them~(namely, the cycles), making the series of Equation~\ref{eq:bfs_non_conv} divergent. However, since the objective is not to know the number of walks between each pair of vertices, but simply to know which vertices are reachable from each other (i.e. there is at least a path between them), this technical problem can be solved by multiplying the adjacency matrix $\mathbf{J}(H)$ by a scalar $\rho_H$, i.e.
\begin{equation}
\label{eq:J_rho}
\mathbf{J}_\rho(H) = \rho_H \ \mathbf{J}(H),
\end{equation}
\noindent
such that
\begin{equation}
\label{eq:rho_H}
0 < \rho_H < \frac{1}{ \rho(\mathbf{J}(H))},
\end{equation}
where $\rho(\mathbf{J}(H))$ is the spectral radius of the matrix $ \mathbf{J}(H)$. As a consequence, the spectral radius of the matrix $\mathbf{J}_\rho(H) < 1$. 
This also results that Equation~\ref{eq:bfs_non_conv} constructed with the matrix $\mathbf{J}_\rho(H)$ converges. Since the convergence of the series is assured, Equation~\ref{eq:bfs_non_conv} can be re-expressed as
\begin{equation}
\label{eq:bfs_conv}
\mathbf{B} = (I - \mathbf{J}_\rho(H))^{-1}.
\end{equation}
The matrix $\mathbf{B}$ defined in Equation~\ref{eq:bfs_conv} has the property that, for any given composite vertex $\mathbf{v} \in \mathbb{V}(H)$,  the row $D(\mathbf{v})$ of $\mathbf{B}$ has non-zero entries in every column that corresponds to a composite vertex $\mathbf{u} \in \mathbb{V}(H)$, such that $\mathbf{u}$ is reachable from~$\mathbf{v}$. Hence, for a given composite vertex $\mathbf{v}$, the row $D(\mathbf{v})$ corresponds to the result of a BFS started at that composite vertex.

Since we now have an algebraic formulation for the BFS, given by either Equation~\ref{eq:bfs_non_conv} or Equation~\ref{eq:bfs_conv}, we can show that the order in which the sub-determination and the BFS are executed matters. Note that the equation 
\begin{equation}
\label{eq:subfist}
\sum_{i=0}^{\infty}  \left( \mathbf{M}_\zeta(H) \ \mathbf{J}_\rho(H) \ \mathbf{M}_\zeta(H)^T  \right)^i
\end{equation}
represents the case where the sub-determination is calculated first and the BFS is calculated afterward while the equation 
\begin{equation}
\label{eq:subaft}
\mathbf{M}_\zeta(H) \ \left( \sum_{i=0}^{\infty} \mathbf{J}_\rho(H)^i \right) \ \mathbf{M}_\zeta(H)^T
\end{equation}
represents the case where the BFS is calculated first and the sub-determination afterward.
Note that in general the result obtained by Equation~\ref{eq:subfist} is different from the result obtained from Equation~\ref{eq:subaft}. To see that this claim holds, note that an arbitrary power $k$ of the matrix $\ \mathbf{M}_\zeta(H)$ $\ \mathbf{J}_\rho(H) \ \mathbf{M}_\zeta(H)^T$ is given by
\begin{equation}
\label{eq:Jzeta}
\underbrace{\mathbf{M}_\zeta(H)\mathbf{J}_\rho(H) \mathbf{M}_\zeta(H)^T  \ \mathbf{M}_\zeta(H) \mathbf{J}_\rho(H) \mathbf{M}_\zeta(H)^T  \dots}_\text{k times}, 
\end{equation}
where  $ \left( \mathbf{M}_\zeta(H) \ \mathbf{J}_\rho(H) \ \mathbf{M}_\zeta(H)^T  \right)$ is multiplied $k$ times. Note, however, that
\begin{equation}
\label{eq:neqMZ}
\mathbf{M}_\zeta(H)^T  \ \mathbf{M}_\zeta(H) \neq \mathbf{I}_n,
\end{equation}
since $\mathbf{M}_\zeta(H) \in \mathbb{R}^{n_\zeta \times n}$ is a rectangular matrix and $n_\zeta~<~n$, so that the rank of the matrix $\mathbf{M}_\zeta(H)^T  \ \mathbf{M}_\zeta(H)$ is less or equal to $n_\zeta$, while the rank of the identity $\mathbf{I}_n$ is $n > n_\zeta$. Since Inequality~\ref{eq:neqMZ} holds, it follows that our claim holds.

From the fact that Equation~\ref{eq:subfist} and Equation~\ref{eq:subaft} are not equivalent, we claim that the results obtained from Equation~\ref{eq:subaft} do not consider spurious paths that can be created by sub-determination. Note that the term $\left( \sum_{i=0}^{\infty} \mathbf{J}_\rho(H)^i \right)$ present in Equation~\ref{eq:subaft} is the BFS calculated for the original (i.e. not sub-determined) MAG $H$. Therefore, it follows that no spurious path is considered in this BFS since no sub-determination was done. As this BFS was calculated with no spurious paths, the result can now be sub-determined for calculating shortest-path centralities without considering unwanted artifacts that may lead to incorrect centrality results.

\section{Centrality in MAGs}
\label{sec:cent_mag}
In graph theory and network analysis, a centrality can be understood as an indicator of the relative importance of the vertices or edges on the graph under analysis. Formally, it follows that for a given graph $G = (V,E)$, 
a vertex centrality $C_v$ can be seen as a function from the set of vertices of a graph to a set of nonnegative real numbers, i.e.
\begin{equation}
\label{eq:vtx_cent}
C_v: V(G) \to \mathbb{R}^+ \bigcup \{0\},
\end{equation}
where $V(G)$ is the vertex set of the graph $G$.
Note that Function~\ref{eq:vtx_cent} induces a linear order relation upon its domains, which can, in turn, be understood as a centrality notion. 
The adopted centrality function should then properly reflect the kind of vertex relative importance to be expressed by the centrality notion.
Similarly, edge centralities may be defined to reflect the relative importance of an edge.

For instance, the vertex betweenness centrality~\cite{Freeman1977} is defined as  
\begin{equation}
\label{eq:vtx_bet}
C_B(v) = \sum_{s,t \in V(G)} \frac{\sigma(s,t/v)}{\sigma(s,t)},
\end{equation}
where $V(G)$ is the the vertex set of the graph $G$, $s,t \in V(G)$ are vertices of the graph G, $v \in V(G)$ is the vertex for which the centrality is calculated, $\sigma(s,t)$ is the number of shortest paths connecting vertices $s$ and $t$, and $\sigma(s,t/v)$ is the number of shortest paths from $s$ to $t$ which pass through vertex $v$. If calculated for each $v \in V(G)$, Equation~\ref{eq:vtx_bet} can be seen as a possible implementation of Function~\ref{eq:vtx_cent}.

\subsection{Extending the centrality notion to MAGs}
\label{subsec:mag_cent}
The evaluation of edge centralities in a MAG can be done in a straightforward way using the MAG's composite vertex representation. Since this representation is a traditional directed graph and it carries all the edges of the MAG, thus preserving their topological properties~(i.e., their adjacency properties), it follows that edge centralities can be computed using the same methods applied to traditional directed graphs.

However, the evaluation of vertex centralities in MAGs is more complex than in a traditional graph. The first difference to be considered is that a MAG can be sub-determined and at the limit any aspect element can become a vertex. Nevertheless, the composite vertices representation of a MAG, which is shown to be isomorphic to the MAG~\cite{Wehmuth2016}, is a directed graph. Hence, vertex centralities in a MAG can be defined in terms of the composite vertices present on the MAG's composite vertices representation, as well as, in terms of any sub-determination of the composite vertices.

\subsection{Composite vertex centralities}
\label{subsec:cent_comp}
Formally, a composite vertex centrality can be seen as a function that goes from the set of composite vertices to the set of nonnegative real numbers, i.e.
\begin{equation}
\mathbb{C}_v: \mathbb{V}(H) \to \mathbb{R}^+ \bigcup \{0\},
\end{equation}
where $\mathbb{V}(H)$ is the set of all composite vertices in $H$~(see Eq.~\ref{eq:VH}).

As the MAG composite vertex representation is a directed graph in which the vertices are the MAG's composite vertices, it follows that any centrality form known for directed graphs can be directly applied to calculate MAG composite vertex centralities.
For instance, the composite vertex betweenness centrality can be defined as
\begin{align}
\label{exp:bet}
\mathbb{C}_v^B: \mathbb{V}(H) & \to \mathbb{R}^+ \bigcup \{0\} \\
\mathbf{v} & \mapsto \sum_{\mathbf{s},\mathbf{t} \in \mathbb{V}(H)} \frac{\sigma(\mathbf{s},\mathbf{t}/\mathbf{v})}{\sigma(\mathbf{s},\mathbf{t})}, \notag
\end{align}
where $\mathbf{v},\mathbf{s},\mathbf{t} \in \mathbb{V}(H)$ are composite vertices of the MAG $H$, $\sigma(\mathbf{s},\mathbf{t})$ is the number of shortest paths from $\mathbf{s}$ to $\mathbf{t}$, and $\sigma(\mathbf{s},\mathbf{t}/\mathbf{v})$ is the number of shortest paths from $\mathbf{s}$ to $\mathbf{t}$ which pass through $\mathbf{v}$.

\subsection{Sub-determined vertex centralities}
\label{subsec:cent_sub}
The definition of centralities for sub-determined vertices on a MAG is quite similar to the centrality definition for composite vertices, i.e.
\begin{equation}
\mathbb{C}_{v_{\zeta}}:\mathbb{V}_{\zeta}(H) \to  \mathbb{R}^+ \bigcup \{0\},
\end{equation}
where $\mathbb{V}_{\zeta}(H)$ is the set of sub-determined vertices of the MAG $H$.
However, the computation of such sub-determined centrality may require distinct algorithms as the computation of composite vertices centralities. For instance, consider the sub-determined betweenness centrality, defined as
\begin{align}
\label{exp:subbet}
\mathbb{C}_{v_{\zeta}}^B: \mathbb{V}_{\zeta}(H) & \to \mathbb{R}^+ \bigcup \{0\} \\
\mathbf{v}_{\zeta} & \mapsto \sum_{\mathbf{s}_{\zeta},\mathbf{t}_{\zeta} \in \mathbb{V}_{\zeta}(H)} \frac{\sigma(\mathbf{s}_{\zeta},\mathbf{t}_{\zeta}/\mathbf{v}_{\zeta})}{\sigma(\mathbf{s}_{\zeta},\mathbf{t}_{\zeta})}, \notag
\end{align}
where $\mathbf{v}_{\zeta},\mathbf{s}_{\zeta},\mathbf{t}_{\zeta} \in \mathbb{V}_{\zeta}(H)$ are sub-determined composite vertices of the MAG $H$ by a given sub-determination $\zeta$, $\sigma(\mathbf{s}_{\zeta},\mathbf{t}_{\zeta})$ is the number of shortest paths from $\mathbf{s}_{\zeta}$ to $\mathbf{t}_{\zeta}$, and $\sigma(\mathbf{s}_{\zeta},\mathbf{t}_{\zeta}/\mathbf{v}_{\zeta})$ is the number of shortest paths from $\mathbf{s}_{\zeta}$ to $\mathbf{t}_{\zeta}$ which pass through $\mathbf{v}_{\zeta}$.

The shortest paths between a given pair of sub-determined vertices $\mathbf{s}_{\zeta},\mathbf{t}_{\zeta} \in \mathbb{V}_{\zeta}(H)$ cannot, however, be computed with the traditional algorithm used for composite vertices. 
Actually, the computation of $\sigma(\mathbf{s}_{\zeta},\mathbf{t}_{\zeta}/\mathbf{v}_{\zeta})$ and,
therefore, all centralities that are defined in terms of shortest paths have to be computed with algorithms constructed based upon the sub-determined shortest paths algorithms discussed in
Section~\ref{sec:def_mag}.

\subsection{Single aspect centralities}
\label{subsec:cent_asp}
Formally, for a given MAG $H = (A,E)$, a single aspect centrality can be seen as a function that takes each element from a given aspect to a nonnegative real number, i.e.
\begin{equation}
\mathbb{C}_a: A(H)[j] \to \mathbb{R}^+ \bigcup \{0\},
\end{equation}
where $A(H)[j]$ is the $j$-th aspect of the MAG $H$.

We highlight that a single aspect centrality is a particular case of sub-determined centralities, since the set $A(H)[j]$ can be obtained by a sub-determination $\zeta_j$ in which the $j$-th entry of the sub-determination tuple is $1$ and all other entries are $0$. Therefore, a single aspect centrality is a particular case of sub-determined centrality, and can then be computed using the same algorithms used for sub-determined centralities. Hence, given any sub-determination 
$\zeta_1$ that preserves a single aspect, the sub-determined betweenness centrality $\mathbb{C}_{v_{\zeta_1}}^B$ as defined in Expression~\ref{exp:subbet} is a single aspect betweenness centrality.
Another example of a single aspect centrality is degree centrality. In this case, the centrality function simply takes each aspect element to the number of edges incident to it. 
In particular, for any MAG that has an aspect designated as time, a time centrality is simply an aspect centrality for this particular aspect~\cite{Costa2015}. 



\section{Algorithms}
\label{sec:Algs}
In this section, we show with a betweenness centrality example how centrality algorithms for MAGs can be derived from well-known algorithms for directed graphs.

We start by presenting Algorithm~\ref{alg:Bet}, which evaluates betweenness centrality for all composite vertices of a given MAG, as defined by Expression~\ref{exp:bet}. We remark that this is just an implementation of the Brandes algorithm~\cite{Brandes2001a, Brandes2008}, which for unweighted sparse networks has as time complexity $O(nm)$ and as space complexity $O(n+m)$, where $n$ is the number of composite vertices and $m$ the number of edges. Note that, for practical purposes, multidimensional complex networks are in general sparse. The original Brandes algorithm can be used directly to evaluate the centralities of composite vertices of a MAG since the composite vertices of a MAG form a directed graph. Therefore, all of the known properties of the Brandes algorithm for directed graphs apply for MAGs when evaluating betweenness centralities for composite vertices. Since this algorithm is well-known, we have included its complete description in Algorithm~\ref{alg:Bet},   mainly to serve as a basis for comparison with the sub-determined betweenness centrality algorithm we present in Algorithm~\ref{alg:Sub}.

\begin{algorithm}[ht]
\scriptsize
\DontPrintSemicolon
	\SetKwData{Left}{left}\SetKwData{This}{this}\SetKwData{Up}{up} 
	\SetKwFunction{Union}{Union}\SetKwFunction{FindCompress}{FindCompress} 
	\SetKwInOut{Input}{input}\SetKwInOut{Output}{output}
	\SetKwFunction{algo}{algo}
        \SetKwProg{myalg}{BetweennessCentrality($H, \tau(H)$)}{}{}
	\Input{$H, \tau(H)$\tcp*{MAG H and companion tuple}}
	\Output{$\mathbb{C}_v^B$\tcp*{Betweenness centrality vector}}
	\BlankLine
	\myalg{} 	{
		$n \gets |V(H)|$ \tcp*{Number of composite vertices of the MAG}
		$\mathbb{C}_v^B \gets$ vector of $n$ integers, all 0 \\
		\For {$\mathbf{s}$ in $V(H)$\tcp*{Single Source shortest path (SSSP) problem} }{ 
			$S \gets$ empty stack  \\
			$P \gets$ a vector of $n$ empty lists \\
			$\sigma \gets$ vector of $n$ integers for counting shortest paths \\
			$\sigma[\mathbf{s}] \gets 1$\tcp*{Each vertex has 1 shortest path to itself}
			$d \gets$ vector of $n$ values $-1$ for distances \\
			$d[\mathbf{s}] = 0$\tcp*{Distance from $s$ to itself is $0$}
			$Q \gets$ queue containing $\mathbf{s}$ \\
			\While{$Q$ is not empty \tcp*{Composite vertices BFS}}{
				dequeue $\mathbf{v} \gets Q$ \\
				push $\mathbf{v} \to S$ \\
				\For {each composite vertex $\mathbf{w}$ such that $(\mathbf{v},\mathbf{w}) \in E(H)$}{
					\If {$d[\mathbf{w}] = -1$ \tcp*{If w found for the first time}} {
					 	enqueue $\mathbf{w}  \to Q$\\
						$d[\mathbf{w}] \gets d[\mathbf{v}] + 1$\tcp*{Distance $\mathbf{s} \to \mathbf{w}$}
					}
					\If {$d[\mathbf{w}] = d[\mathbf{v}] + 1$\tcp*{If there is a shortest path to $\mathbf{w}$ via $\mathbf{v}$}} {
						$\sigma[\mathbf{w}] \gets \sigma[\mathbf{w}]  + \sigma[\mathbf{v}]$\tcp*{Counts shortest paths} 
						append $\mathbf{v} \to P[\mathbf{w}] $\tcp*{$\mathbf{v}$ is predecessor of $\mathbf{w}$}
					}
				}
			}
			$\delta \gets$ vector of $n$ integers, all 0\\
			\While{$S$ is not empty} {
				pop $\mathbf{w} \gets S$\\
				\For {$\mathbf{v}$ in $P[w]$ \tcp*{For all predecessors of $\mathbf{w}$}} {
					$\delta[\mathbf{v}] \gets \delta[\mathbf{v}] + (\sigma[\mathbf{v}]  \mathbf{/} \sigma[\mathbf{w}]) \times (1 + \delta[\mathbf{w}])$ 
				}
				\If{$\mathbf{w} != \mathbf{s}$} {
					$\mathbb{C}_v^B[\mathbf{w}] \gets \mathbb{C}_v^B[\mathbf{w}] + \delta[\mathbf{w}]$
				}
			}
		}
	}
	\Return{$\mathbb{C}_v^B$}\;
\caption{Betweenness centrality for composite vertices.}
\label{alg:Bet}
\end{algorithm} 

When considering the sub-determined form of betweenness 
centrality, defined by Expression~\ref{exp:subbet}, we first have to take into account that calculating betweenness centrality of a sub-determined MAG is not the same as calculating the sub-determined betweenness centrality of a MAG. This is a consequence of the fact that composite vertices that are not connected in a MAG may be connected on a sub-determined form of this MAG, meaning that the sub-determination process may create paths on the sub-determined MAG that have no real correspondence to paths existing on the original MAG. As an illustrative example of this, we present Figures~\ref{fig_a}~and~\ref{fig_b}.  The MAG $R$ shown at Figure~\ref{fig_a} can be seen either as a multilayer or a time-varying network. It can be seen that starting from vertex $1$ it is not possible to reach vertex $3$. However, in Figure~\ref{fig_b}, which is a sub-determined form of the MAG $R$, there is a path connecting vertex $1$ to vertex $3$. 

As a consequence of this spurious connectivity, shortest paths that should not exist can be present on a sub-determined form of a MAG. A complete description of this undesired effect can be found at~\cite{Wehmuth2017}. In order to prevent this to influence the results of the evaluation of betweenness centrality, we propose a sub-determined algorithm, which uses the original (non sub-determined) MAG to evaluate centrality. 

The sub-determined betweenness centrality, presented in Algorithm~\ref{alg:Sub}, is based upon the original Brandes algorithm. Algorithm~\ref{alg:Sub} evaluates sub-determined betweenness centrality (i.e. betweenness centrality of sub-determined vertices) without considering the spurious paths potentially generated by MAG sub-determination.

\begin{algorithm}[t!]
\scriptsize
\DontPrintSemicolon
	\SetKwData{Left}{left}\SetKwData{This}{this}\SetKwData{Up}{up} 
	\SetKwFunction{Union}{Union}\SetKwFunction{FindCompress}{FindCompress} 
	\SetKwInOut{Input}{input}\SetKwInOut{Output}{output}
	\SetKwFunction{algo}{algo}
        \SetKwProg{myalg}{SubBetweennessCentrality($H, \tau(H),\zeta$)}{}{}
	\Input{$H, \tau(H), \zeta$\tcp*{MAG H, companion tuple, and sub-determination}}
	\Output{$\mathbb{C}_{v_{\zeta}}^B$\tcp*{Betweenness centrality vector}}
	\BlankLine
	\myalg{} 	{
		$n \gets |V(H)|$ \tcp*{Number of composite vertices of the MAG}
		$n_\zeta \gets |V_\zeta(H)|$\tcp*{Number of composite vertices of the sub-determined MAG}
		$\mathbb{C}_{v_{\zeta}}^B \gets$ vector of $n_\zeta$ integers, all 0 \\
		\For {$\mathbf{s_\zeta}$ in $V_\zeta(H)$\tcp*{Single Source shortest path (SSSP) problem} }{ 
			$color \gets$ vector of $n$ integers, all $0$ \\
			$color_\zeta \gets$ vector of $n_\zeta$ integers, all $0$ \\
			$S \gets$ empty stack  \\
			$P \gets$ a vector of $n_\zeta$ empty lists \\
			$\sigma \gets$ vector of $n_\zeta$ integers for counting shortest paths \\
			$\sigma[\mathbf{s_\zeta}] \gets 1$\tcp*{Each vertex has 1 shortest path to itself}
			$d \gets$ vector of $n_\zeta$ values $-1$ for distances \\
			$d[\mathbf{s_\zeta}] = 0$\tcp*{Distance from $s$ to itself is $0$}
			$Q \gets$ empty queue \\
			\For {$\mathbf{i}$ in $V_\zeta(H)$, where $s_\zeta == i_\zeta$ \tcp*{For all vertices with same sub-det. as $\mathbf{s}$} } {
				$color[\mathbf{i}] \gets $ BLACK \\
				enqueue $\mathbf{i} \to Q$ \tcp*{All vertices equivalent to $\mathbf{s}$}
			} 
			\While{$Q$ is not empty \tcp*{Sub-determined vertices BFS}}{
				dequeue $\mathbf{v} \gets Q$ \\
				$\mathbf{v_\zeta} \gets \mathbf{v}$ sub-determined by $\zeta$ \\
				\If {$color_\zeta[\mathbf{v_\zeta}]$ is WHITE}{
					$color_\zeta[\mathbf{v_\zeta}] \gets$ BLACK \\
					push $\mathbf{v_\zeta} \to S$ \\
				} 
				\For {each composite vertex $\mathbf{w}$ such that $(\mathbf{v},\mathbf{w}) \in E(H)$}{
					\If {$color[\mathbf{w}]$ is WHITE}{
						$color[\mathbf{w}] \gets$ BLACK \\
						enqueue $\mathbf{w}  \to Q$\\
					}
					$\mathbf{w_\zeta} \gets \mathbf{w}$ sub-determined by $\zeta$ \\
					\If {$d[\mathbf{w_\zeta}] = -1$ \tcp*{If $\mathbf{w_\zeta}$ found for the first time}} {
						$d[\mathbf{w_\zeta}] \gets d[\mathbf{v_\zeta}] + 1$\tcp*{Distance $\mathbf{s_\zeta} \to \mathbf{w_\zeta}$}
					}
					\If {$d[\mathbf{w_\zeta}] = d[\mathbf{v_\zeta}] + 1$\tcp*{If there is a shortest path to $\mathbf{w_\zeta}$ via $\mathbf{v_\zeta}$}} {
						$\sigma[\mathbf{w_\zeta}] \gets \sigma[\mathbf{w_\zeta}]  + \sigma[\mathbf{v_\zeta}]$\tcp*{Counts shortest paths} 
						append $\mathbf{v_\zeta} \to P[\mathbf{w_\zeta}] $\tcp*{$\mathbf{v_\zeta}$ is predecessor of $\mathbf{w_\zeta}$}
					}
				}
			}
			$\delta \gets$ vector of $n_\zeta$ integers, all 0\\
			\While{$S$ is not empty} {
				pop $\mathbf{w_\zeta} \gets S$\\
				\For {$\mathbf{v_\zeta}$ in $P[w_\zeta]$ \tcp*{For all predecessors of $\mathbf{w_\zeta}$}} {
					$\delta[\mathbf{v_\zeta}] \gets \delta[\mathbf{v_\zeta}] + (\sigma[\mathbf{v_\zeta}]  \mathbf{/} \sigma[\mathbf{w_\zeta}]) \times (1 + \delta[\mathbf{w_\zeta}])$ 
				}
				\If{$\mathbf{w_\zeta} != \mathbf{s_\zeta}$} {
					$\mathbb{C}_{v_{\zeta}}^B[\mathbf{w_\zeta}] \gets \mathbb{C}_{v_{\zeta}}^B[\mathbf{w_\zeta}] + \delta[\mathbf{w_\zeta}]$
				}
			}
		}
	}
	\Return{$\mathbb{C}_{v_{\zeta}}^B$}\;
\caption{Sub-determined betweenness centrality for MAGs.}
\label{alg:Sub}
\end{algorithm} 

The key differences between Algorithm~\ref{alg:Sub} and the traditional Brandes algorithm is that in Algorithm~\ref{alg:Sub} both the sub-determined and the full (non sub-determined) MAG are taken into account. Actually, the Breadth-First Search~(BFS) part of the algorithm is done upon the full MAG, while the paths are built upon the sub-determined vertices. By doing this, we have that only the paths that exist on the full MAG will be taken into account for calculating the centrality, whose result is expressed in terms of sub-determined vertices.

By comparing both the traditional and the sub-determined algorithms, it can be seen that the main differences are at lines 3, 8, 16 to 20, and 30 to 33. In Algorithm~\ref{alg:Sub}, for each sub-determined vertex of the MAG, a BFS is conducted using the original (full) MAG and shortest paths are computed for each distinct sub-determined vertex found by the BFS. Further, it is important to note that given a sub-determined vertex $\mathbf{s_\zeta}$, instead of starting from a single vertex, the BFS starts from all vertices whose sub-determined form is equal to $\mathbf{s_\zeta}$. This is the equivalent of starting a BFS on a sub-determined MAG. However, since the BFS is conducted upon the not sub-determined MAG, it follows that the BFS respects the reachability found on the full MAG and, therefore, it \emph{disregards the spurious paths} created by sub-determining the MAG. Further, since the functional difference between Algorithm~\ref{alg:Sub} and the traditional betweenness centrality algorithm is that Algorithm~\ref{alg:Sub} uses the sub-determined BFS algorithm proposed in~\cite{Wehmuth2017}, it follows that Algorithm~\ref{alg:Sub} has the same properties of the traditional Brandes algorithm. In particular, since the BFS problem is done over the full MAG, and if we consider that $n_\zeta$ (the number of $\zeta$ sub-determined vertices) is of the same order of $n$ (the number of composite vertices), it follows that Algorithm~\ref{alg:Sub} can be computed in $O(nm)$ time and requires $O(n + m)$ space, given that this algorithm is for unweighted MAGs~(note that $m$ is the number of edges). It is worth noting that this adaptation can be implemented for any other shortest path based algorithm, such as closeness, stress and other centralities of this kind.

As an illustrative example, we remark that the betweenness centrality calculated for the sub-determined MAG~$R$ shown in Figure~\ref{fig_b} is [0, 1, 0] while the sub-determined betweenness centrality for the MAG $R$ shown in Figure~\ref{fig_a} is [0, 0, 0]. From this example, it can be seen that the betweenness centrality (Algorithm~\ref{alg:Bet}) considers the presence of a path from vertex $1$ to $3$ in Figure~\ref{fig_b}, making the centrality of vertex $2$ to have value~$1$. In the case of sub-determined betweenness centrality (Algorithm~\ref{alg:Sub}) of the MAG $R$ shown in Figure~\ref{fig_a}, the result is [0, 0, 0], from which it can be seen that no path from vertex $1$ to $3$ is considered in this case.

\section{Illustrative case studies}
\label{sec:app}

In this section, we present a set of case studies designed to show, in an experimental approach, the differences between the outcomes of centrality measures when using classical centrality algorithms on previously sub-determined MAGs or when  using sub-determined algorithms applied on (non-sub-determined) MAGs. The observed differences illustrate the impact on centrality analysis by mistakenly considering spurious paths in aggregated views of multidimensional networks as compared with avoiding such spurious paths in the aggregation process, as proposed in this paper, before the centrality analysis.


\subsection{Real-world and synthetic networks}

In our case studies, we consider two real-world multidimensional networks as well as synthetic ER (Erd\"{o}s-R\'{e}nyi) random MAGs: 

\begin{itemize}

\item MOSAR network -- MOSAR (Mastering hOSpital Antimicrobial Resistance and its spread) is a scientific collaboration project~\cite{MOSAR} that comprises several medical, biochemistry, and computing research institutions. The MOSAR project focuses on antimicrobial-resistant bacteria (AMRB) transmission dynamics in high-risk environments, such as intensive care units and surgical centres. The adopted dataset consists of the records of in-person contacts (from physicians, nurses, staff members, and patients) in a certain medical ward for a period of two weeks~(between 12 am of July 25, 2009 and 12am of August 08, 2009). Each one of the 160 volunteers who participated in the study was equipped with a RFID device that detected the presence of another devices within a small distance~(about one meter). Device identification was unique and was always associated with the same person. Every 30 seconds during the two-weeks period, each device registered the list of all devices (nodes) that were within its coverage area in order to establish the arrangement of the contacts among them (edges). We have then modeled the MOSAR dataset as a order~2 MAG representing a TVG with 160 participants (elements in the first aspect) and 40320 time instants~(elements in the second aspect). Contacts are represented by non-oriented edges, i.e., the edge between any two persons is represented by a directed edge and its reciprocal. With spurious paths, the characteristic path length and diameter of the MOSAR network are~2.39 and~5, respectively. Disregarding spurious paths, the characteristic path length and diameter of the MOSAR network are~2.64 and~9, respectively.

\item Brazilian domestic air transportation network -- The second real-world dataset we use in our experiments is the air transportation network from Brazil, as published by the Brazilian National Civic Aviation Agency~(ANAC) in May, 12th 2017. We model this  dataset as an order~3 MAG to represent the multidimensional network equivalent to a multilayer time-varying network we detail as follows. The first aspect consists of 110~elements corresponding to the airports. The second aspect consists of the multilayer part in which each layer involves the air flight network of each airline companies. In this second aspect, there are 14~elements as 2~layers are used for each of the 7~airline companies present in the dataset. The third aspect is composed of 7057 elements~(i.e., time instants). This arrangement results in a MAG with 48339~composite vertices and 66195~edges. This is a weighted network. The weights in the edges represent the time duration of flights, boardings, landings, and the waiting time between flights. As previously indicated, each airline company uses 2~layers: one layer of the flights and the other layer for the waiting times between flights. The layer with waiting times is where we put the edge to make the weekly cycle. With spurious paths, the characteristic path length and diameter of this network are~2.39 and~5, respectively. Disregarding spurious paths, the characteristic path length and diameter are~2.47 and~7, respectively.

\item Synthetic ER random MAGs -- We have generated 2000 synthetic ER random MAGs, where 1000 are order~2 MAGs and the other 1000 are order~3 MAGs. The order~2 MAGs may represent time-varying graphs~(TVGs) while the order~3 MAGs may represent higher order networks, such multilayer time-varying networks. Each sample for both kinds of synthetic ER random MAGs has 10000 composite vertices connected through 42586 edges. The difference is that these 10000 composite vertices are divided in the order~2 MAG as 1000 elements in the first aspect and 10 elements in the second aspect; while in the order~3 MAG as 1000 elements in the first aspect, 2 elements in the second aspect, and 5 elements in the third aspect. The characteristic path length in ER random networks is $O(\log n)$ and the diameter $l \simeq \log n / \log \bar{k}$, where $\bar{k}=2m/n$ is the average degree~\cite{Barabasi-book20016}. This results in a characteristic path length in the order of 4~hops and an expected value for the diameter of about 4.30~hops synthetic ER random MAGs.

\end{itemize}

\subsection{Methodology for the experimental study}

For all the analyses we conduct in the presented networks, we sub-determine the MAG representing the multidimensional network into a single aspect and then we evaluate the betweenness and the closeness centralities. For instance, for each random MAG, the results correspond to the centrality of the first aspect elements. In its turn, the centrality we analyze in the sub-determined MOSAR network refers to each person of the original face-to-face network~(shown as node IDs in Section~\ref{subsec:mosar}). Finally, for the Brazilian air transportation network, the single aspect puts forward the centrality of the airports~(shown in airport codes in Section~\ref{subsec:air}). 

Moreover, we compare the results from centrality metrics computed by the classical approach and by our approach, developed to evaluate sub-determined centralities on MAG, thus avoiding spurious paths. We actually compare the centrality rankings generated by both cases and we analyze the position changes when the two rankings are compared. More specifically, to compare the rankings, we use the Ranking-Biased Overlap~(RBO)~\cite{Webber2010}, which is a ranking similarity measure. In contrast to the classic Kendall coefficient to compare rankings, RBO allows us to evaluate the similarity between rankings by applying weights (i.e. distinct importance) to the positions of the considered rankings. For example, the top positions of the rank (and their value) may have higher importance than positions far from the beginning of the rank. 
Note that RBO allows us to control how to weight rank positions. For example, we are able to fix $x$ percent of the weight to the first $n$ elements of the rank. Also note that RBO measures similarity, not distance (i.e., RBO is not a metric). However, RBO can trivially be turned into a distance measure, denominated rank-biased distance~(RBD), where RBD = 1 - RBO, as proved in~\cite{Webber2010}.
In particular, for all experiments presented in this section, RBO and RBD will be set to assign $85\%$ of the total weight to the top $10\%$ positions of the rankings.

In short, as we previously stated, our goal is to show that, in the case of shortest path based centralities, the classic centrality algorithms may fail due to the issues resulting from the presence of spurious paths caused by the aggregation (and sub-determination).

\subsection{Evaluating MOSAR -- a face-to-face human contact network}
\label{subsec:mosar}

\begin{figure}[htbp]
        \centering
        \includegraphics[width=0.4\textwidth]{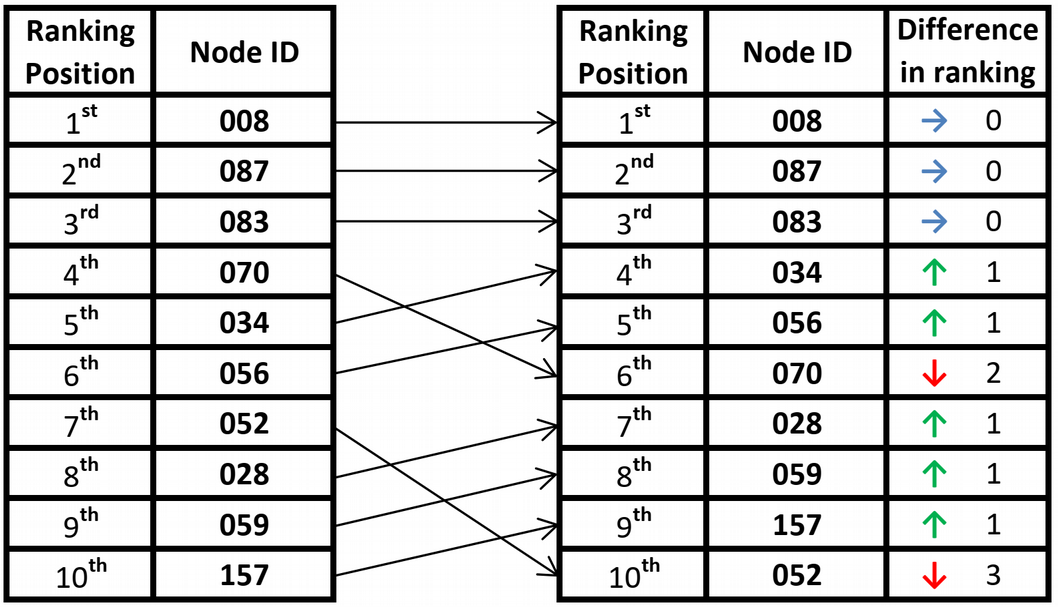}
        \caption{Rankings for the betweenness centrality in the MOSAR network when (a)~spurious paths may be present~(left ranking); and (b)~spurious paths are avoided~(right ranking).}
        \label{fig:mosar_bet}
\end{figure}

\begin{figure}[htbp]
        \centering
        \includegraphics[width=0.4\textwidth]{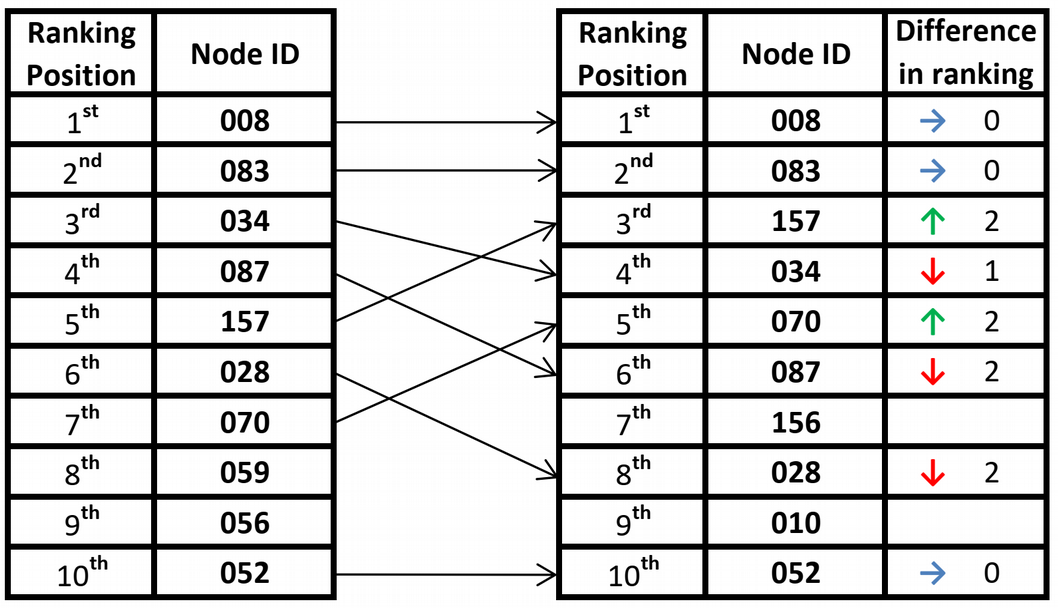}
        \caption{Rankings for the closeness centrality in the MOSAR network when (a)~spurious paths may be present~(left ranking); and (b)~spurious paths are avoided~(right ranking).}
        \label{fig:mosar_close}
\end{figure}

In this experiment, we evaluate the difference obtained when applying the proper algorithms for sub-determined results and classical sub-determined algorithms on the MOSAR network.
Figures~\ref{fig:mosar_bet}~and~\ref{fig:mosar_close} show the top 10 vertices for, respectively,  the betweenness and the closeness centralities for both approaches (i.e., the classical sub-determined algorithms and the use of proper algorithms for sub-determined results). In this sense, we compare the centrality rankings considering the presence of spurious paths (or not). In both figures, for both network centralities, we clearly note a difference between the position of the top 10 vertices, a result of the spurious paths. In fact, according to Figure~\ref{fig:mosar_bet}, which shows the resutls for betweenness centrality, the two rankings differ from the third position. The rankings for closeness centrality, shown in Figure~\ref{fig:mosar_close}, diverge from the fourth position on the ranking. Moreover, in the case of Figure~\ref{fig:mosar_bet}, we also note the presence of elements that do not figure on both top~$10\%$ rankings.

When comparing the top  $10\%$ of central nodes, the rankings of betweenness centrality present a significant RBD distance. In this case, the RBD distance between the top $10\%$ of the ranking is $0.188$, while the RBD distance for the closeness centrality is slightly smaller~($0.119$).
In short, spurious paths incur in a considerable misleading assessment of node centrality.

\subsection{Evaluating the Brazilian domestic air transportation network}
\label{subsec:air}

\begin{figure}[htbp]
        \centering
        \includegraphics[width=0.4\textwidth]{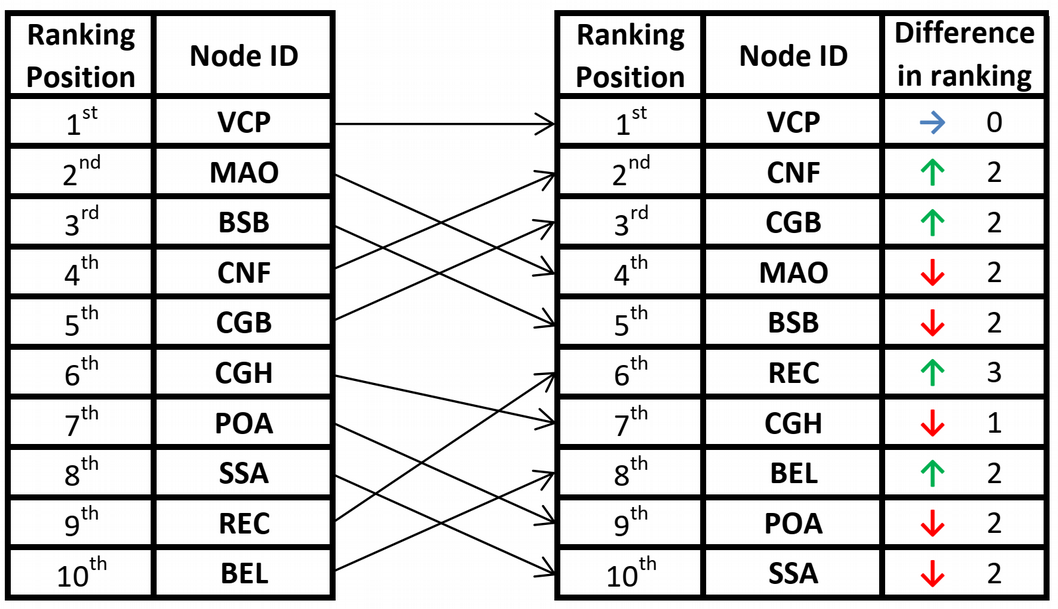}
        \caption{Rankings for the betweenness centrality in the Brazilian air transportation network when (a)~spurious paths may be present~(left ranking); and (b)~spurious paths are avoided~(right ranking).}
        \label{fig:malha_bet}
\end{figure}

\begin{figure}[htbp]
        \centering
        \includegraphics[width=0.4\textwidth]{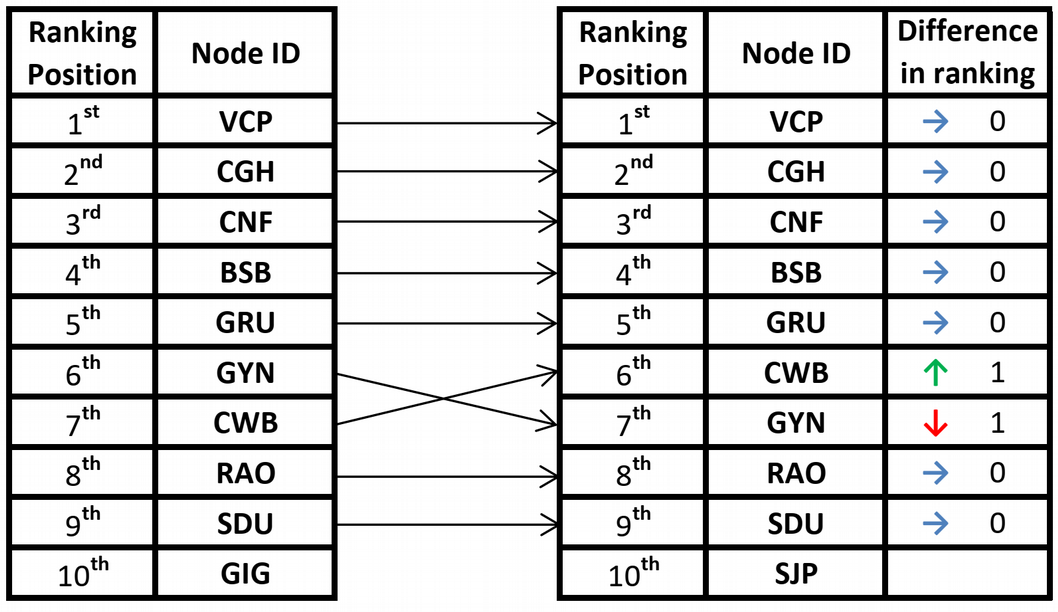}
        \caption{Rankings for the closeness centrality in the Brazilian air transportation network when (a)~spurious paths may be present~(left ranking); and (b)~spurious paths are avoided~(right ranking).}
        \label{fig:malha_close}
\end{figure}

We evaluate the betweenness and closeness centralities of the Brazilian domestic air transportation network. 
Figures~\ref{fig:malha_bet} and~\ref{fig:malha_close} graphically highlight the ranking differences. According to Figure~\ref{fig:malha_bet}, the difference between the two betweenness centrality rankings is important. For the betweenness, we note important pair exchanges in the ranking positions, starting from the second position of the ranking. On the other hand, as shown in Figure~\ref{fig:malha_close}, the differences between both closeness centrality rankings is smaller. We only note one pair exchaging their relative position and, a pair of distinct airports in the 10th ranking position. This similarity between both rankings is expected once the closeness centrality algorithm calculates a mean of the distances from each vertex, therefore reducing the influence of spurious paths.

In this evaluation, using the RBO, we have assigned 85\% of the ranking weight to the top 10\% of the ranking (i.e., 11 vertices). In this case, we observe an RBO correlation to the pair of betweenness centrality rankings of 0.787. The RBO correlation to the pair of closeness centrality rankings is of 0.956. In other words, the distance between betweenness centrality rankings is of $0.213$, while the RBD distance for the closeness centrality rankings is considerably smaller~(i.e. 0.044).

\subsection{Evaluating synthetic  ER  random  MAGs}

In this section, we present the correlation and distances obtained for betweenness and closeness centralities, considering the algorithms suppressing spurious paths and the classic algorithms that do not suppress spurious paths. We first evaluate the $1000$ synthetic ER random order~2 MAGs~(equivalent to time-varying graphs -- TVGs) and then we evaluate the $1000$ synthetic ER random order~3 MAGs. Unless indicated otherwise, in this section, we consider that 85\% of the RBO and RBD weight is assigned upon the first 10\% of the rankings.

Tables~\ref{tab:bet_rboTVG}~and~\ref{tab:close_rboTVG} summarize the RBO and RBD values we observe for the betweenness and the closeness centralities when comparing rankings of TVG (order~2 MAG) networks with spurious paths and without spurious paths.
First, we clearly note that spurious paths impose a relevant effect on the centrality metrics we analyze.  For example, for the betweenness centratlity~(Table~\ref{tab:bet_rboTVG}), the mean RBD value between both rankings is~0.386. Moreover, the lowest difference observed is of 0.175. The RBD values for the closeness centrality~(Table~\ref{tab:close_rboTVG}) are as high as those for  betweenness centrality 
In this case, the mean RBD value is superior to 0.6, while the minimum RBD is about 0.37. In this sense, analyzing path-related centralities using the classical approach to TVG networks may lead to misconceptions about the centrality assessment of the network.

\begin{table}[htbp]
\caption{Betweenness centrality: RBO and RBD values for random TVGs (order~2 MAGs).}
\label{tab:bet_rboTVG}
\centering
\begin{tabular}{|c|c|c|}
\hline
\textbf{TVG - Betweenness}     & \textbf{RBO} & \textbf{RBD} \\ \hline
\textbf{Minimum} & 0.404        & 0.175        \\ \hline
\textbf{Maximum} & 0.825        & 0.596        \\ \hline
\textbf{Mean}  & 0.614        & 0.386        \\ \hline
\textbf{Standard Deviation}       & 0.072        & 0.072        \\ \hline
\end{tabular}
\end{table}

\begin{table}[htbp]
\caption{Closeness centrality: RBO and RBD values for random TVGs (order~2 MAGs).}
\label{tab:close_rboTVG}
\centering
\begin{tabular}{|c|c|c|}
\hline
\textbf{TVG - Closeness}          & \textbf{RBO} & \textbf{RBD} \\ \hline
\textbf{Minimum} & 0.273        & 0.369        \\ \hline
\textbf{Maximum} & 0.631        & 0.727        \\ \hline
\textbf{Mean}  & 0.397        & 0.603        \\ \hline
\textbf{Standard Deviation}       & 0.066        & 0.066        \\ \hline
\end{tabular}
\end{table}

We also observe high RBD values in the case of the random order~3 MAGs for both betweeness and closeness centralities, as shown in Tables~\ref{tab:bet_rboMAG}~and~\ref{tab:close_rboMAG}.
For example, the mean RBD for betweenness and closeness centralities are superior to 0.39 and 0.6, respectively. Again, this leads to the conclusion that the path-based centrality computation in random order~3 MAGs are also significantly impacted by spurious paths induced through the sub-determination process.

\begin{table}[htbp]
\caption{Betweenness Centrality: RBO and RBD values for random order~3 MAGs.}
\label{tab:bet_rboMAG}
\centering
\begin{tabular}{|c|c|c|}
\hline
\textbf{MAG - Betweenness}       & \textbf{RBO} & \textbf{RBD} \\ \hline
\textbf{Minimum} & 0.409        & 0.173        \\ \hline
\textbf{Maximum} & 0.827        & 0.591        \\ \hline
\textbf{Mean}  & 0.608        & 0.392        \\ \hline
\textbf{Standard Deviation}       & 0.072        & 0.072        \\ \hline
\end{tabular}
\end{table}

\begin{table}[!h!]
\caption{Closeness Centrality: RBO and RBD values for random order~3 MAGs.}
\label{tab:close_rboMAG}
\centering
\begin{tabular}{|c|c|c|}
\hline
\textbf{MAG - Closeness}         & \textbf{RBO} & \textbf{RBD} \\ \hline
\textbf{Minimum} & 0.267        & 0.368        \\ \hline
\textbf{Maximum} & 0.632        & 0.733        \\ \hline
\textbf{Mean}  & 0.398        & 0.602        \\ \hline
\textbf{Standard Deviation}       & 0.064        & 0.064        \\ \hline
\end{tabular}
\end{table}


\section{Conclusion}
\label{sec:rem}

In this paper, we study shortest-path based centralities in multidimensional (high order) complex networks. Usually, the assessment of node centralities in these networks involves the aggregation of different dimensions to then perform the computation of centrality metrics. Nevertheless, this aggregation process may create spurious paths on the aggregated view of the network, in particular when time-varying graphs are involved. Computing shortest-path based centralities, such as betweennees or closeness, in the presence of spurious paths may then mislead the intended network centrality analysis. 

To the best of our knowledge, we propose the first method to avoid taking into account spurious paths at the aggregation process in multidimensional complex networks. This assures the aggregated view of the higher order network to be free of spurious paths, thus allowing an accurate shortest-path based centrality computation in multidimensional complex networks. 
To achieve this result, we first represent multidimensional complex networks by means of MultiAspect Graphs~(MAGs),  an abstraction able to represent time-varying, multilayer, combined time-varying multilayer, or even higher order graphs. 

We have evaluated both, the traditional approach to calculate centrality metrics in multidimensional networks disregarding the potential existence of spurious paths and our proposal that avoids the generation of spurious paths at the aggregation process. In other words, we compare betweenness and closeness centralities for the case where the algorithms include the spurious paths generated by sub-determination and for the case where the algorithms are constructed to properly avoid this issue. Such a case study is interesting because it shows impact of properly avoiding the issue of spurious paths on the network centrality computation.

Finally, as an additional contribution, we make a python implementation of the algorithms discussed in this paper publicly available.\footnote{\url{http://github.com/wehmuthklaus/MAG_Algorithms}}


\section*{Acknowledgment}

This work was partially funded by CAPES, CNPq, FAPEMIG, FAPESP, and FAPERJ. Authors also acknowledge the INCT in Data Science -- INCT-CiD.

\bibliographystyle{plain}
\bibliography{library} 

\end{document}